\documentclass[11pt]{article} \usepackage{hyperref} \pdfoutput=1
\begin{document} \title{Wetting Splashing}
\author{Peichun (Amy) Tsai, Maurice Hendrix, Remko Dijkstra, and Detlef Lohse  \\ \\\vspace{6pt} Department of Applied Physics, \\ University of Twente, The Netherlands}
\maketitle
\begin{abstract} 
We present fluid dynamics videos illustrating wetting splashing--produced by water drop impact onto hydrophobic microstructures at high impact velocity ($\sim 3$ ms$^{-1}$).  The substrate consists of regular and transparent microtextures in square or hexagonal lattice, yielding a high contact angle of $\sim 150 ^\circ$.  Our high speed top-or-bottom view recordings through the transparent surface shed light on the solid-liquid-air interfaces at impact.  Despite the superhydrophobicity of the latticed micropillars (of a periodicity of $10~\mu$m), water droplet wets a certain central area and moreover entraps an air bubble beneath the droplet. Besides the central wet area, lamella surf on air splashing outward at high impinging velocity. The effects of micropatterns and air pressure on the impact outcome are also examined.  We show that microscopic boundary condition, imposed by the solid texture, profoundly influences the macroscopic flow dynamics upon superhydrophobic surfaces at high impinging velocity. In addition, the intervening air between the liquid and the solid plays a crucial role in directional splash, which can be eliminated by a reduced air pressure.  
\end{abstract}
\end{document}